\documentclass[letterpaper, 10 pt, conference,fleqn]{ieeeconf}  
\usepackage{graphicx}
\usepackage{amsmath}     
\usepackage{xcolor}

\IEEEoverridecommandlockouts 
\title{Gain-loss ratio of storing intermediate data from workflows 
}

\author{Debasish Chakroborti
\\debasish.chakroborti@usask.ca
}

\begin{document}

\maketitle
\thispagestyle{empty}
\pagestyle{empty}

\section{Introduction}

Workflow management systems(WfMS) are now being commonly used for monitoring systems, increasing productivity, reducing costs, error recovery, improving information exchange, and introducing reusability in various scientific, engineering, and business organizations. In Big-data analytics, when a large volume of heterogeneous data needs to be processed with various mechanisms, workflow management systems (WfMS) should be considered for efficiency. A WfMS can be built on local as well as distributed and parallel computing environments such as clusters, grids, and clouds.
In a workflow management system, intermediate states play a vital role in keeping track of a complete execution flow for provenance trace and other purposes. However, storing data (results) from all the intermediate states of a workflow is not a good idea as it requires large memory. In a workflow-based system, datasets are usually large, and thus, their intermediate states may also become considerably large. While there are existing studies and techniques for storing all of the intermediate states in a system, only a few studies discuss candidate policy of intermediate states without a specific solution rather than saving all. Here we describe the researches \cite{Chakroborti2020} \cite{8622351} \cite{10.1145/3457145} that deal with a recommendation system of workflows for storing intermediate states. This recommendation process is designed to serve the purpose of reusability considering frequently used data usage patterns with co-relation among modules in a workflow. The term pipeline is also used commonly to process various sequential modules in a system instead of workflow, and in this paper, we will use both terms interchangeably.

\begin{figure}[thpb]
      \centering
      
      \includegraphics[width=0.5\textwidth]{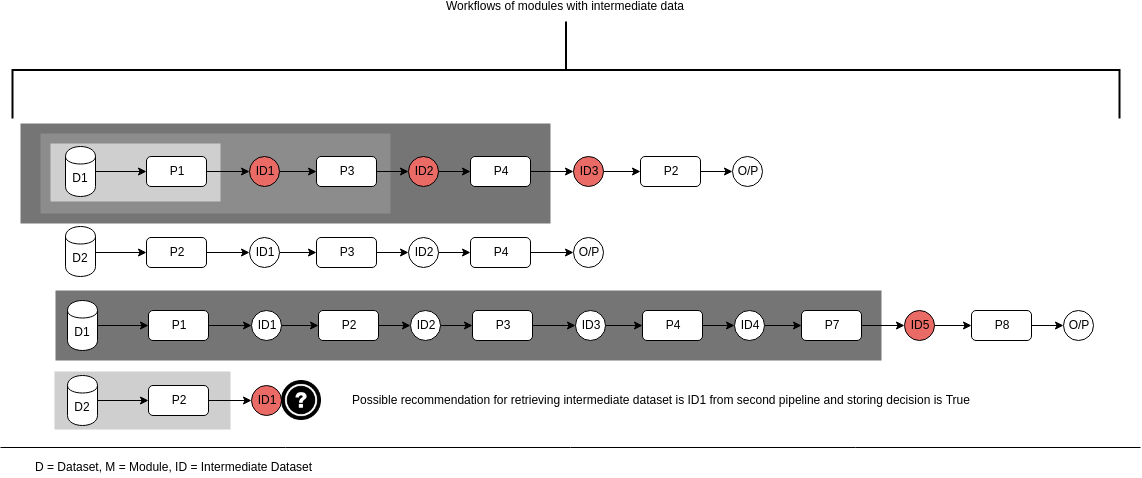}
      \caption{Automatically suggesting pipeline retrieving and storing points for intermediate  data}
      \label{workflows}
\end{figure}

Storing intermediate data and use of those data for re-usability and other purposes in a system are frequent tasks. In a system of workflow management, if we store each piece of intermediate states without knowing their correlation with other modules or workflows, after a certain period, this will not be cost-effective, and there will not be any possibility to introduce a re-usability mechanism. Also, appropriate intermediate data searching is a time-consuming task as different heterogeneous data are processed with various modules in different workflows. Thus an optimal data suggestion mechanism is also required in a system for the re-usability. Selecting a candidate set to store intermediate states and searching intermediate datasets for a specific purpose need time and prior experience. In such cases, association rules for the past usage pattern can give us valuable information to make a recommendation system for both data selection and store, and it would be beneficial for both researcher/user. Such a technique can reduce a huge amount of time for pipeline making and help both researcher and user experiment with different types of data.

\textbf{Proposed techniques.} The current studies propose and investigate recommendation techniques that rely on association rule mining for recommending which intermediate state result(s) from a pipeline under progress should be stored for reuse in the future - proposed technique RISP (Recommending Intermediate States from Pipelines) in \cite{8622351}. The technique makes association rules between data and processing modules by examining the pipeline usage history of a WfMS and analyzing these association rules to identify which intermediate state results from the pipeline under progress have a high possibility of being reused in the future. Such intermediate state results with high possibilities are recommended for storing. RISP not only provides recommendations regarding storing intermediate state results but also suggests reusing already stored intermediate state results by analyzing data and module usage patterns of the pipeline that a user is currently building.

The study proposes and investigates a recommendation system for both storing and retrieving processes of intermediate datasets, which uses the association rule mining technique for a workflow when a researcher/user is currently developing to accomplish a specific job. Association technique can use past usage information to find correlations among workflows or modules and can help to define the suggestion system.

At each workflow generation and execution process, usage history is automatically stored, and the association mining technique is applied on all of the previous workflow information to learn a pattern of usage of modules with datasets in workflows. In the future, when a user attempts to build a workflow, our recommendation system provides useful information to select intermediate datasets as well as recommends for which data need to be stored as intermediate states.

Support and confidence two measures of association rule have been used to rank the dataset for the recommendation. These two measures help RISP determine which of the existing datasets have a high probability of being selected as intermediate datasets in the pipeline under progress and which datasets are needed to be stored in the system. Association rule and support and confidence have been discussed in section III.

In Figure \ref{workflows}, two datasets: D1 D2, are used in four pipelines. Each pipeline has some processing modules. For instance, the first pipeline that processes dataset D1 has four processing modules P1, P2, P3, and P4. Also, each pipeline has some intermediate states of a raw dataset. In the first pipeline, three intermediate datasets, ID1, ID2, and ID3, are generated after modules P1, P3, and P4. Suppose the system supports the re-usability of processed data, and a researcher needs to decide at each point of intermediate data for both reusing and storing from/to the system. Each decision point depends on subsets of some modules and datasets. For example, the third workflow's red-marked intermediate dataset point depends on dataset D1 and modules P1 to P7, and 4th pipeline's ID1 depends on dataset D2 and modules P2. When a user builds the fourth workflow, that time decision for retrieving at ID1 can be assertive as the subset (D2, P2) already occurs in the second pipeline. Also, storing decisions can be true as this subset is frequent compared to others. For providing such suggestions, the recommendation system derives and analyzes association rules among modules in workflows in the system.

\section{Mechanism for Determining Association Rules from Pipelines}

Let us assume that we have the usage history of workflow pipelines for a certain period of time. By analyzing the history, RISP will determine which intermediate state results should be stored for which data set. Let us consider the following four pipelines for describing the idea.

~\\$Pipeline1:  D1: P1-P3-P4-P2$ \\
$Pipeline2:    D2: P2-P3-P4$ \\
$Pipeline3:    D1: P1-P2-P3-P4-P7-P8$ \\
$Pipeline4:    D2: P2-P4-P5-P7$ \\

We now consider the first pipeline, which consists of four processes (P1, P3, P4, and P2 sequentially) uses the dataset D1. Let us define a sub-pipeline that processes data by applying some processes and provides an intermediate result. From the first pipeline, we can define three such sub-pipelines. These are: (p1), (p1, p3), and (p1, p3, p4). Each of these sub-pipelines produces some intermediate result. We need to decide whether it will be beneficial to store the intermediate result from any one of these sub-pipeline considering dataset D1. By considering the three sub-pipelines from the first pipeline and the dataset D1 (as it was used in the first pipeline), we define the following three-item sets:

~\\$[D1, (p1)]                 [D1, (p1, p3)]                 [D1, (p1, p3, p4)]$\\

In the same way, we also determine item sets from the other three pipelines. All the 12 item sets from all four pipelines are listed below.

~\\$[D1, (p1)]                 [D1, (p1, p3)]                 [D1, (p1, p3, p4)]$\\

~\\$[D2, (p2)]                 [D2, (p2, p3)]$\\

~\\$[D1, (p1)]                 [D1, (p1, p2)]                 [D1, (p1, p2, p3)]$                 
~\\$[D1, (p1, p2, p3, p4)]                 [D1, (p1, p2, p3, p4,p7)]$\\

~\\$[D2, (p2)]                 [D2, (p2, p4)]                 [D2, (p2, P4, P5)]$\\

We determine an association rule from each of these item sets. For example, we determine the following association rule from the item set [D1, (p1, p3, p4)].

~\\$D1 => (p1, p3, p4)$\\

The meaning of such an association rule is that if some uses the dataset D1, then there is a possibility that she will sequentially execute the three processes p1, p3, and p4 starting from p1. This conditional probability is called the confidence of the association rule, and we determine this confidence from the support values of the constituent entities in the following way.

~\\$Confidence (D1 => (p1, p3, p4)) = Support (D1 => (p1, p3, p4)) / Support (D1)$\\

Support is the number of times an item occurred in the entire usage history. Our usage history consists of the 12 item sets from the four pipelines. From this usage history, we get the following values of supports for $(D1 => (p1, p3, p4)) and (D1).$.

~\\$Support (D1 => (p1, p3, p4)) = Support (D1, (p1, p3, p4)) = 2$

~\\$Support (D1) = 7$

~\\$Thus, Confidence (D1 => (p1, p3, p4)) = 2/7$\\

Similarly, we determine the confidences for the distinct association rules from all the item sets. We then rank these association rules on the basis of their confidence values. Association rules with high confidence values provide us opportunities for storing intermediate results. For example, the confidence values of the eight distinct association rules from the 12 item sets are listed here. The rules are ranked according to their confidence values.

~\\$Confidence (D2 => (p2)) = 2/5$\\
$Confidence (D1 => (p1)) = 2/8$\\
$Confidence (D1 => (p1, p3)) = 2/8$\\
$Confidence (D1 => (p1, p3, p4)) = 2/8$\\
$Confidence (D2 => (p2, p3)) = 1/5$\\
$Confidence (D2 => (p2, p4)) = 1/5$\\
$Confidence (D2 => (p2, p4, p5)) = 1/5$\\
$Confidence (D1 => (p1, p3, p4, p7)) = 1/8$\\

We see that the rule $D2 => (p2)$ has the highest confidence. However, it suggests us storing intermediate rules from a sub-pipeline that consists of one process only. If we consider the next three rules, we see that they have the same confidence value. In this case, we can select the rule $D1 => (p1, p3, p4)$ because it includes the consequents of the two previous rules in sequential order. The other association rules have confidence values. From such an example, it will be beneficial if we can store the intermediate result produced by the sub-pipeline (p1, p3, p4) considering the dataset D1.

\section{Related Work}

D. Yuan et al. \cite{Yuan2011On-demandSystems} state that many scientific workflows use intermediate states to reuse or share data in a system. But most of the systems select candidate sets to store intermediate states with a manual investigation. In their work, making a dependency graph, they propose an algorithm for a minimum cost of saving intermediate states. Their trade-off model for computation and storage with cost model shows effectiveness over various experiments.

E. H. Shennan et al. \cite{ShennanUsingKnowledge} illustrate the importance of intermediate states to reuse and share knowledge in clinical tools. 

E. M. Hart et al. \cite{HartTenStorage} define various rules of the data storage for a large amount of data. While giving the directions, they emphasize repository-based solutions for data storage with Meta-data where data sharing should be reflected. Also, raw and intermediate datasets should be distinguishable by proper documentation.   


S. Woodman et al. \cite{Woodman2015WorkflowProvenance} point that for workflow provenance, full traces of data and ownership are essential. But storing all of the full tracks using intermediate states is costly, and as a solution, they propose a cost model where various candidate sets are generated with a binary selection fashion, and the lowest cost candidate set of the intermediate states can be stored for efficiency.   

J. Han et al. \cite{Han2000MiningGeneration} demonstrate a technique of frequent patterns determination without generating all costly candidate sets. In their study, they propose a tree-based mining method for frequent transaction analysis in a database system.


J. A. Thompson et al. \cite{ThompsonADNA} illustrate various tools to select candidate sets in RNA analysis using a workflow. 

M. A. Wolters \cite{Wolters2015AProblems} proposes an algorithm for a fixed size candidate set selection, named kofnGA for Applications using GA.

M. Lozano et al. \cite{Lozano2016AProblem} propose a meta-heuristic-based solution for the MSG problem using a genetic algorithm, where a technique based on knapsack problem does the candidate set selection. 

A. Darehshoorzadeh et al. \cite{DarehshoorzadehCandidateRouting} compare four algorithms for network routing when an optimal solution is not necessary to implement in a system, OR Algorithms perform better than traditional scheduling is reported in the study. A system can perform fast for frequent changes to the routing list using an opportunistic model is stated in their work.  

K. Krawiec et al. \cite{KrawiecMetaheuristicRepair} explore the possibility to use a repair technique for a meta-heuristic-based solution when a candidate solution is not a feasible one.

J. R. Dunphy et al. \cite{Dunphy2003IntelligentSpacesb} propose a preprocessing technique of candidate set before applying their genetic algorithm to reduce the problem space. 

S. Ghosh et al. \cite{Ghosh2010MiningAlgorithm} suggest a genetic algorithm for mining frequent itemsets rather than using a traditional costly items mining algorithm.

\section{Background}
\label{background}

Association rules have been heavily used in software engineering research and practice for performing impact analysis tasks. For example, when a programmer attempts to make changes to a program entity (such as files, classes, methods), association rules can help us identify which other entities might get impacted because of this change. We apply the association rule mining technique in our research for automatically suggesting a module to add in a partially complete pipeline. We define an association rule in the following way.

\textbf{Association Rule.}

An association rule \cite{AgrawalMiningDatabases} is an expression of the form $X => Y$ where X is the antecedent and Y is the consequent. Each of X and Y is a set of one or more program entities. The meaning of such a rule in our context is that if X gets changed in a particular commit operation, Y also has the tendency of being changed in that commit.

\textbf{Support and Confidence.}
As defined by  Zimmermann et al. \cite{ZimmermannMiningChanges}, \emph{support is the number of commits in which an entity or a group of entities changed together}. We consider an example of two entities \emph{$E_1$} and \emph{$E_2$}.  If \emph{$E_1$} and \emph{$E_2$} have ever changed together, we can assume two association rules, $E_1 => E_2$ and $E_2 => E_1$.
Suppose, \emph{$E_1$} was changed in four commits: 2, 5, 6, and 10 and \emph{$E_2$} was changed in six commits: 4, 6, 7, 8, 10, and 13. So, \textit{support($E_1$) = 4} and \textit{support($E_2$) = 6}. However, \textit{support($E_1$, $E_2$) = 2}, because \emph{$E_1$} and \emph{$E_2$} co-changed (changed together) in two commits: 6 and 10.  Support of a rule is determined in the following way.

\begin{equation}
\mathit{support}(X=>Y)=\mathit{support}(X, Y)
\end{equation}

Here, $(X, Y)$ is the union of $X$ and $Y$, and so $\mathit{support}(X=> Y) = \mathit{support}(Y => X)$. For the above example, $\mathit{support} (E_1 => E_2) = \mathit{support}(E_2 => E_1) = \mathit{support}(E_1, E_2) = 2$.

\emph{Confidence of an association rule, $X => Y$, determines the probability that $Y$ will change in a commit operation provided that $X$ changed in that commit operation}. We determine the confidence of $X => Y$ in the following way.

\begin{equation}
\mathit{confidence}(X => Y) = \mathit{support}(X, Y) / \mathit{support}(X)
\end{equation}

For the example, \emph{confidence ($E_1 => E_2$) = support($E_1$, $E_2$) / support($E_1$) = 2 / 4 = 0.5} and \emph{confidence($E_2 => E_1$) = 2 / 6 = 0.33}. In our research, we derive association rules from pipelines and investigate those for providing suggestions to complete a pipeline under progress.


\section{Recommending intermediate states for storing and retrieving using association rules}
\label{proposal}

\subsection{Determining association rules from pipelines}
We derive association rules among modules and datasets by analyzing existing pipelines in the following way. Let us consider the first pipeline in Fig. \ref{workflows}. This pipeline consists of four modules: P1, P3, P4, P2 and three intermediate states ID1, ID2, ID3 and one raw dataset. If we consider every possible subset of consecutive modules in this pipeline to process raw data, we get the following subset: (p1), (p1, p3), and (p1, p3, p4). From each of these pairs, we can determine one association rule. For example, from the first pair consisting of modules P1 and dataset D1, we derive the association rules $(D1 => (P1))$. Such an association rule means that if a subset of a pipeline process dataset D1, then there is a possibility that in the future, the same subset will process D1. In the association rule $(D1 => (P1))$, D1 is called the \textbf{antecedent} and (P1) is called the \textbf{consequent}. From the three pairs in the first pipeline, we can determine three association rules. From all the three complete pipelines in Fig. \ref{workflows}, we determine the  following  eight distinct association rules: $(D2 => (p2)), (D1 => (p1)), (D1 => (p1, p3)), (D1 => (p1, p3, p4)), (D2 => (p2, p3)), (D2 => (p2, p4)), (D2 => (p2, p4, p5)), (D1 => (p1, p3, p4, p7))$

\subsection{Determining the supports and confidences of the  association rules obtained from the pipelines}
After determining all the distinct association rules from all the pipelines of given usage history, we determine the support and confidence of each of the rules. For example, we will now determine the support and confidence of the association rule $(D1 => (P1))$. 

\begin{itemize}
    \item Support $(D1 => (P1))$ = 2, because in two complete pipelines (the first and the third) in Fig. \ref{workflows} we see that the module P1 appears just after input data set D1. Here, we should note that the traditional way of computing support value for an association rule will assume the same support value for two association rules: $(D1 => (P1, X))$ and $(D1 => (X, P1))$ [X is assumed as another element in the subset]. While such consideration is reasonable for co-changing software entities where the frequency of co-change matters but the order in which they were changed together does not matter, it is not a reasonable consideration for our research. In our case, the order of appearance of the two modules also matters. The output that we will get by sequentially applying P1 and X modules on a piece of data will be different from the output that we will obtain by sequentially applying X and P1. Thus, for example in Fig. \ref{workflows}, Support $(D1 => (X,P1))$ = 0, because P1 did not appear just after X in any pipeline. 
    
    \item
    Confidence $(D1 => (P1))$ = Support $(D1 => (P1))$ / Support (D1) = 2/8 (the highest value of confidence). Here, Support (D1) is the number of times it appears in the complete pipelines. Such a confidence means that if D1 is added in an image processing pipeline, then according to the history there is a very high possibility that (P1) will be added just after it. We can also calculate that Confidence $(D1 => (p1, p3))$ = Support $(D1 => (p1, p3))$ / Support (D1) = 2/8, and Confidence $(D1 => (p1, p3, p4))$ = 2/8. As Confidences $(D1 => (P1))$, $(D1 => (p1, p3))$ and $(D1 => (p1, p3, p4))$ are higher than other Confidences, our recommendation system will prioritize these three points over others when automatically suggesting the saving points as well as possible dataset to be used in the fourth pipeline (i.e., the partially complete one).
\end{itemize}

\begin{figure}
    \includegraphics[width=0.49\textwidth]{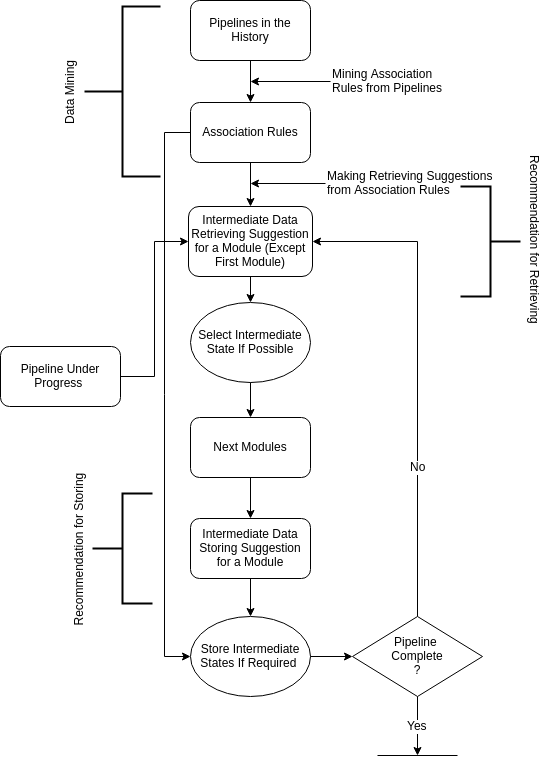}    
    \caption{Steps for suggesting intermediate states in a pipeline}
    \label{system}
\end{figure}

\subsection{Automatically recommending intermediate data by analyzing association rules}
\label{recommendingmodules}
Let us assume that a user is going to build a pipeline for performing a particular task in a workflow management system (WMS). As soon as she adds a module in her pipeline, our automatic recommendation system performs the following steps sequentially in order to suggest available intermediate data for skipping some processing modules and storing intermediate data in her pipeline.

\begin{itemize}
    \item \textbf{Step 1:} It collects all the pipelines from the usage history of the WMS. The usage history may contain multiple occurrences of the same pipeline. Our recommendation system considers all the occurrences. Let us denote the set of all pipelines by P.
    
    \item \textbf{Step 2:} For each of the pipelines in the set P, the system determines all possible association rules of modules. As it proceeds by selecting and analyzing each pipeline in P, it keeps track of which datasets as well as which association rules appeared how many times in history. From this information, it determines the supports of datasets and association rules. After getting the support, a system can determine the confidence of the association rules. It finally generates a set of distinct association rules with respective support and confidence values from the workflow usage history of the workflow management system.
    
    \item \textbf{Step 3:} From the association rules obtained from the previous step, the system selects those rules where the antecedent is the same dataset of the pipeline under progress. The consequents of these selected association rules are considered as the suggestions for the intermediate states to skip some modules in the partially completed pipeline if previously stored. Our recommendation system also ranks these suggested modules on the basis of the support and confidence values of their corresponding association rules.
\end{itemize}

After getting data suggestions for the module to add to the pipeline, the user may select one from the list of suggested datasets or the raw one (in this case, all modules need to be executed). After she has selected the data, the system also uses the association rules for storing decisions of an intermediate dataset. The system again performs \textbf{Step 3} to get suggestions for the next module on the basis of the previous module that the user added to her pipeline. In this way, our recommendation system suggests intermediate datasets and storing policy to the user at each stage of building a pipeline. The whole process of intermediate data recommendation has been shown in Fig. \ref{system}. We see that the three boxes just above the first oval and the last box before the last oval represent the outcomes from the three steps we have just described. The figure also shows that the user can repeatedly select her next module's dataset to add to her pipeline by getting suggestions from our recommendation system.

\section{EVALUATION}

This section presents our evaluation with timeframes of workflow generations in our system. Every time there is a new workflow, we can consider a new state of the system, which can be considered a timeframe for the system evaluation. We consider step by step process for these timeframes to evaluate our proposed storing mechanism. At each timeframe, we evaluate the decision we made for previous workflows in the system with nth workflow, so we need to consider 1 to n-1 workflows at every nth workflow generation period. Toward explanation, let us consider after a certain period, we have four pipelines below.\\

\begin{flalign}
\mathit{Pipeline1}(P1=>P3=>P4=>P2) \mathit{\;with\;D1}
\end{flalign}
\begin{flalign}
\mathit{Pipeline2}(P2=>P3=>P4) \mathit{\;with\;D2}
\end{flalign}
\begin{flalign}
\mathit{Pipeline3}(P1=>P2=>P3=>P4=>P7=>P8) \mathit{\;with\;D1}
\end{flalign}
\begin{flalign}
\mathit{Pipeline4}(P2=>P4=>P5=>P7) \mathit{\;with\;D2}
\end{flalign}\\

At the first time, when we have only one pipeline (1), in this case, we can make three subsets $(P1)$,    $(P1, P3)$, $(P1, P3, P4)$ of sequential module's operation for processing the same dataset as described before. Hence at this point, from pipeline-1, we can have a total of three subsets of a pipeline to generate intermediate states, and as a first pipeline, now we do not have any history of pipeline generations. So, in this case, if we calculate confidence for the first pipeline, then we will have similar values for all the subsets, and we need to store all the intermediate stats generated at the endpoint of each subset in a pipeline. Now consider the second pipeline is created in the system, and we need some quantitative values for both decision making and evaluation techniques. First, we need all possible subsets to process raw data, i.e., $(P2)$, $(P2, P3)$ of consecutive modules in the workflow, but this workflow uses a different dataset than the first workflow, and again we can not use history of workflow usage although we have records of the previous workflows. Therefore, again, in this case, we will store all intermediate datasets with a similar confidence value. Now consider the third pipeline where we have five subsets to aim intermediate states' saving point, and this workflow uses the same dataset as a previous workflow which is workflow 1 in this case. Subsets are here $(P1)$,    $(P1, P2)$, $(P1, P2, P3)$, $(P1, P2, P3, P4)$, $(P1, P2, P3, P4, P7)$. For this workflow, now we have a usage history with the same dataset, D1. At this time, we can check how we can use the past usage information and how much it will be helpful for our system. Can we get gain from the past for some of the subsets we have generated in the nth workflow? To calculate the gain, we set a variable for each workflow's subset for gain, loss, and no effect. When a new workflow is generated in the system, subsets are calculated and matched with previous workflows' subsets. If there is a match and previously intermediate states are stored for this subset, then the variable is updated with +1 (gain =1). If there is a match and subset intermediate data are not stored, then the variable is updated with -1 (loss =1). If there is no match, then the variable is updated with 0 (no effect = 1). Following are the effects of 3 workflow generations. 

~\\Associations of workflow 1:\\
$(D1=>(P1))$ GainOrLossOrNoEffect 1\\
$(D1=>(P1, P3))$ GainOrLossOrNoEffect -1\\
$(D1=>(P1, P3, P4))$ GainOrLossOrNoEffect -1\\
$(D1=>(P1, P3, P4, P2))$ GainOrLossOrNoEffect -1\\
Associations of workflow 2:\\
$(D2=>(P2))$ GainOrLossOrNoEffect 0\\
$(D2=>(P2, P3))$ GainOrLossOrNoEffect 0\\
Total Gain 1 and Loss 3 from all the previous workflow's intermediate states storing decision.\\

Now we have to consider some circumstances to evaluate the process. First, when we do not have required past information we store all intermediate states as all the confidences are same. And when we have past information with the same dataset, an intermediate state of a subset with the highest confidence value will be stored. If we calculate the confidence of the subsets' associations at this point, we have the following results.

~\\$(D1=>(P1)) = 2/8 $
~\\$(D1=>(P1, P2)) = 1/8 $
~\\$(D1=>(P1, P2, P3)) = 1/8 $
~\\$(D1=>(P1, P2, P3, P4)) = 1/8 $
~\\$(D1=>(P1, P2, P3, P4, P7)) = 1/8 $
~\\

So our plan for a workflow is to save intermediate states at the point of highest confidence value, as it is frequently used path with the same data and there is a higher possibility
to be used in the future. Now we need to consider workflow 4, where we have three subsets (P 2), (P 2, P 4), (P 2, P 4, P 5). Confidences of the associations are presented below:

~\\$(D2=>(P2)) = 2/5 $\\
$(D2=>(P2, P4)) = 1/5 $\\
$(D2=>(P2, P4, P5)) = 1/5 $\\

Now if we want to explore gain or loss at this point from previous all three workflows, we need association by association information.

~\\Associations of workflow 1:\\
$(D1=>(P1))$ GainOrLossOrNoEffect 1+0\\
$(D1=>(P1, P3))$ GainOrLossOrNoEffect -1+0\\
$(D1=>(P1, P3, P4))$ GainOrLossOrNoEffect -1+0\\
$(D1=>(P1, P3, P4, P2))$ GainOrLossOrNoEffect -1+0\\
Subsets of workflow 2:\\
$(D2=>(P2))$ GainOrLossOrNoEffect 0+1\\
$(D2=>(P2, P3))$ GainOrLossOrNoEffect 0-1\\
Subsets of workflow 3:\\
$(D1=>(P1)) = 2/8 $ GainOrLossOrNoEffect 0+0\\
$(D1=>(P1, P2)) = 1/8 $ GainOrLossOrNoEffect 0+0\\
$(D1=>(P1, P2, P3)) = 1/8 $ GainOrLossOrNoEffect 0+0\\
$(D1=>(P1, P2, P3, P4)) = 1/8 $ GainOrLossOrNoEffect 0+0\\
$(D1=>(P1, P2, P3, P4, P7)) = 1/8 $ GainOrLossOrNoEffect 0+0\\

At this point, the total gain is 2, and the loss is 4. 

Now, if we look at the ratio of gain and loss step by step, we can see that evolving history with the storing procedure can also help us increase the gain and loss ratio.
If we evaluate the full system with this step-by-step process for all workflows in our system, we can see that every time we consider a new workflow, the ratio is increased, which shows that storing intermediate data by the system is useful and more beneficial with more usage history.

\section{CONCLUSIONS}
The aforementioned work investigates the importance of making a candidate set for storing intermediate states rather than saving all of them in a workflow management system. An automated dataset suggestion is also considered to increase the reusability in a pipeline to reduce module execution cost. Investigations on workflows show that there is a high correlation among modules to use datasets, and from this relation, we can make associations for our recommendation system. We also find that using association rules, it is possible to store highly frequent consecutive modules outcome and measure them with gain-loss ratios. We believe that gain-loss ratio analysis is important to understand the reusability, and other factors of workflows need to be investigated with this measure in a WfMS. 
\addtolength{\textheight}{-12cm}  

\bibliographystyle{IEEEtran}
\bibliography{Reference}
\end{document}